\documentclass[twocolumn]{jpsj3}
\usepackage{txfonts}
\usepackage{xcolor}
\usepackage[dvipdfmx]{graphicx}
\usepackage{lscape}
\usepackage{mathcomp} 
\usepackage{textcomp}
\usepackage{cite}

\title{Pressure Effects on CeMnSi: Evolution of Ce-\(4f\) and Mn-\(3d\) Electronic States and Negative Thermal Expansion}
\author{Sae Nishiyama$^1$, Haruka Mima$^1$, Jun-ichi Hayashi$^1$, Keiki Takeda$^1$, Chihiro Sekine$^1$, Yoshiya Uwatoko$^{2, 3}$, Hiroki Takahashi$^4$, Hiroshi Tanida$^5$, and Yukihiro Kawamura$^1$\thanks{y{\_}kawamura@muroran-it.ac.jp}}
\inst{$^1$Muroran Institute of Technology, Muroran, Hokkaido 050-8585, Japan\\ 
$^2$Institute for Solid State Physics, the University of Tokyo, Kashiwa, Chiba 277-8581, Japan\\ 
$^3$Faculty of Science and Engineering, Tokyo City University, Setagaya, Tokyo 158-8557, Japan\\
$^4$College of Humanities and Science, Nihon University, Setagaya, Tokyo 156-8550, Japan\\ $^5$Liberal Arts and Sciences, Toyama Prefectural University, Imizu, Toyama 939-0398, Japan\\}

\abst{
We investigated pressure effects on the nontrivial heavy-fermion antiferromagnet CeMnSi by means of electrical resistivity (\(\rho\)) and powder X-ray diffraction. With increasing pressure, the antiferromagnetic order of Mn (\(T_{\mathrm{N}} \sim 240\) K at ambient pressure) is rapidly suppressed and disappears at \(P_{\mathrm{c}} \sim 1.3\)~GPa. 
Instead, a pressure-induced anomaly appears at \(T_{\mathrm{M}} \sim 97\) K and shifts to higher temperatures with increasing pressure. 
The switching of the Mn magnetic state may reflect a modification of the magnetic symmetry of the system, which could influence the stability of the heavy-fermion state.
In the low-pressure region, non-Fermi-liquid-like behavior characterized by nearly \(T\)-linear resistivity is observed around \(0.7\) GPa. 
In addition, \(\rho\) shows a marked reduction below \(T_{\mathrm{M}}\) and a qualitative change toward more metallic behavior above the structural transition pressure \(P_{\mathrm{s}} \sim 5.7\) GPa. 
At ambient pressure, CeMnSi exhibits negative thermal expansion below \(\sim 40\) K, which is absent in LaMnSi, supporting the formation of a heavy-fermion ground state.
}

\begin{document}
\maketitle

\section{Introduction}

Ce-based intermetallic compounds exhibit a wide variety of correlated-electron phenomena arising from the dual nature of the Ce-\(4f\) electrons, which can behave either as localized magnetic moments or as itinerant heavy quasiparticles depending on external control parameters~\cite{Stewart1984,Gegenwart2008}.
External pressure is one of the most effective tuning parameters, as it enhances the \(c\)--\(f\) hybridization strength and can suppress magnetic order toward a critical pressure at which it disappears. In the vicinity of such a regime, enhanced fluctuations often give rise to unconventional electronic responses, including non-Fermi-liquid-like behavior~\cite{Lohneysen2007}.
In many correlated-electron systems, such electronic instabilities are accompanied by pronounced thermodynamic responses, including anomalous thermal expansion, reflecting the strong coupling between the dominant electronic energy scale and the lattice~\cite{Drotziger2006,Luo2015}.
Thermal expansion and, more generally, lattice responses provide important probes of the coupling between electronic states and the lattice in heavy-fermion systems.

\begin{figure}[tbp]
\begin{center}
\includegraphics[width=1\linewidth]{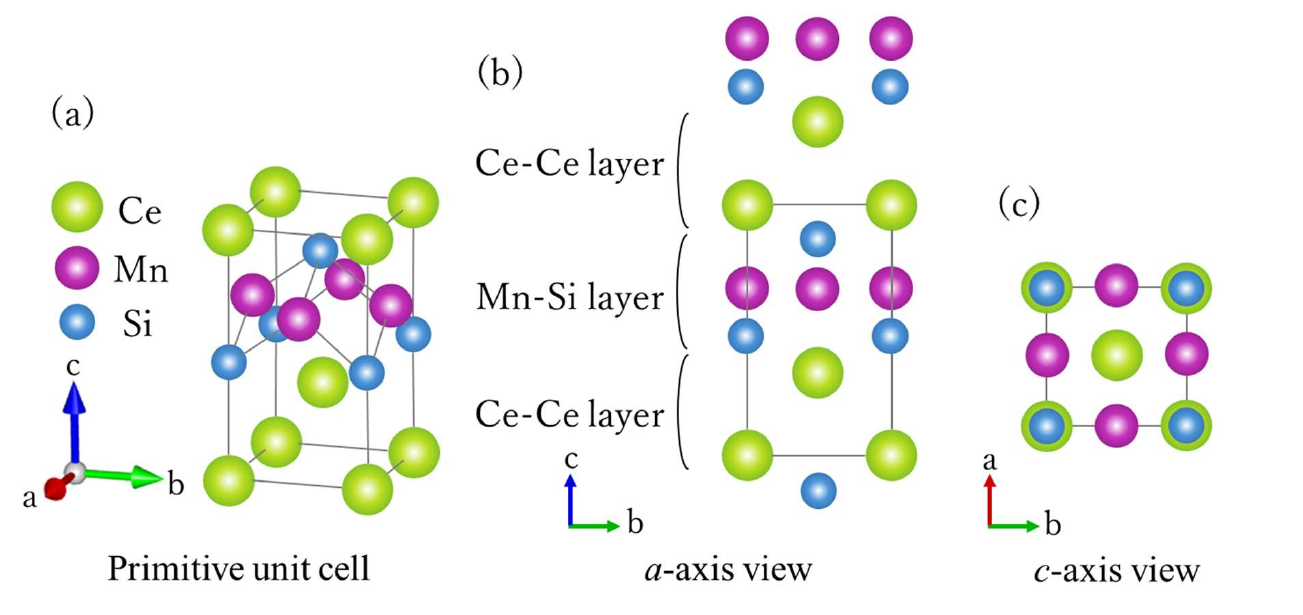}
\caption{(Color online)  Crystal structure of CeMnSi with its layered arrangement, drawn using the VESTA software~\cite{vesta}.
(a) Primitive unit cell; (b) and (c) views along the \(a\) and \(c\) axes, respectively.
The tetragonal CeFeSi-type structure consists of alternating Ce--Ce layers and Mn--Si layers stacked along the \(c\) axis.}
\label{tetra}
\end{center}
\end{figure}

The Ce\(T\)Si (\(T=\) Mn, Co) compounds crystallize in the tetragonal CeFeSi-type structure (space group \(P4/nmm\))~\cite{Bodak1970} and provide a useful platform for examining the interplay between electronic states and lattice responses.
Figure~\ref{tetra} shows the crystal structure of CeMnSi, which consists of alternating Ce--Ce layers and Mn--Si layers stacked along the \(c\) axis. This layered structure provides a structural basis for coupling among localized Ce-\(4f\) electrons, itinerant Mn-\(3d\) electrons, and the lattice.

CeCoSi exhibits an antiferromagnetic ground state accompanied by a hidden-order--like transition,
which has been shown to involve a structural modulation and a lowering of crystal symmetry \cite{Tanida2019,Matsumura2022}.
This finding demonstrates strong coupling between electronic ordering and the lattice in CeCoSi.
In contrast, CeMnSi exhibits a long-range antiferromagnetic order of the Mn-\(3d\) moments at \(T_{\mathrm{N}} \sim 242\)~K,
together with the development of a heavy-fermion state associated with the Ce-\(4f\) electrons at much lower temperatures~\cite{Tanida2023}.
While its magnetic and transport properties have been investigated in detail, the lattice response of CeMnSi at low temperatures
and its evolution under pressure have not yet been clarified.

Furthermore, both CeCoSi and CeMnSi undergo pressure-induced structural phase transitions at high pressures, occurring at the structural transition pressure
\(P_{\mathrm{s}} \sim 4.9\)~GPa for CeCoSi and \(P_{\mathrm{s}} \sim 5.7\)~GPa for CeMnSi~\cite{Kawamura2020,Kawamura2026}.
Notably, the structural transition is absent in non-\(4f\) reference compounds, suggesting that it is closely related to the instability of the Ce-\(4f\) electronic state. 
At the same time, the resulting lattice instability may further influence both Ce-\(4f\) and Mn-\(3d\) electronic states under high pressure. 
It is of interest to investigate pressure effects on both the nontrivial heavy-fermion state and the Mn antiferromagnetic ordering, as well as lattice properties such as thermal expansion.

In this study, we investigate the pressure evolution of the electronic and lattice properties of CeMnSi
by electrical resistivity (\(\rho\)) and powder X-ray diffraction (XRD) measurements over a wide range of temperatures and pressures.
We show that interactions among the Ce-\(4f\) electrons, Mn-\(3d\) electrons, and the lattice give rise
to diverse pressure-induced phases and unconventional physical phenomena.

\section{Experiment}
Single crystals of CeMnSi and LaMnSi were grown using the self-flux method following our previously reported procedure~\cite{Tanida2023}. 
\(\rho\) under pressure was measured using both a cubic-anvil apparatus and a piston--cylinder cell with a standard four-terminal DC method.
The electrical current was applied parallel to the \(ab\) plane of the CeMnSi single crystal. 
Measurements in the low-pressure region up to about 1.5 GPa were carried out using a piston--cylinder cell, 
whereas those at higher pressures were performed using a cubic-anvil apparatus.
A 1:1 mixture of Fluorinert FC70 and FC77 was used as the pressure-transmitting medium. 
The temperature ranges were 2.2--300 K for the cubic-anvil apparatus and 1.3--300 K for the piston--cylinder cell. 
The pressure in the piston--cylinder cell was calibrated using the superconducting transition temperature of Pb~\cite{Pb}.

Powder XRD measurements using synchrotron radiation were performed at BL-18C of the Photon Factory, KEK, Japan. 
The wavelength of the synchrotron radiation was \(\lambda = 0.6200\,\text{\AA}\).
Low-temperature XRD measurements at ambient and high pressures were carried out using a GM-type refrigerator with a Mylar X-ray window. 
For ambient-pressure measurements only, the powdered samples were fixed using Apiezon grease.
High-pressure XRD measurements were performed using a helium gas-driven diamond anvil cell with a CuBe gasket. 
A 4:1 methanol--ethanol mixture was used as the pressure-transmitting medium, and the pressure was determined by ruby fluorescence~\cite{ruby}. 
The measurements were performed over the temperature range 10--300 K at pressures up to about 8 GPa.

\section{Results}
\subsection{Electrical Resistivity under Pressure}  

\begin{figure}[tbp]
\begin{center}
\includegraphics[width=0.8\linewidth]{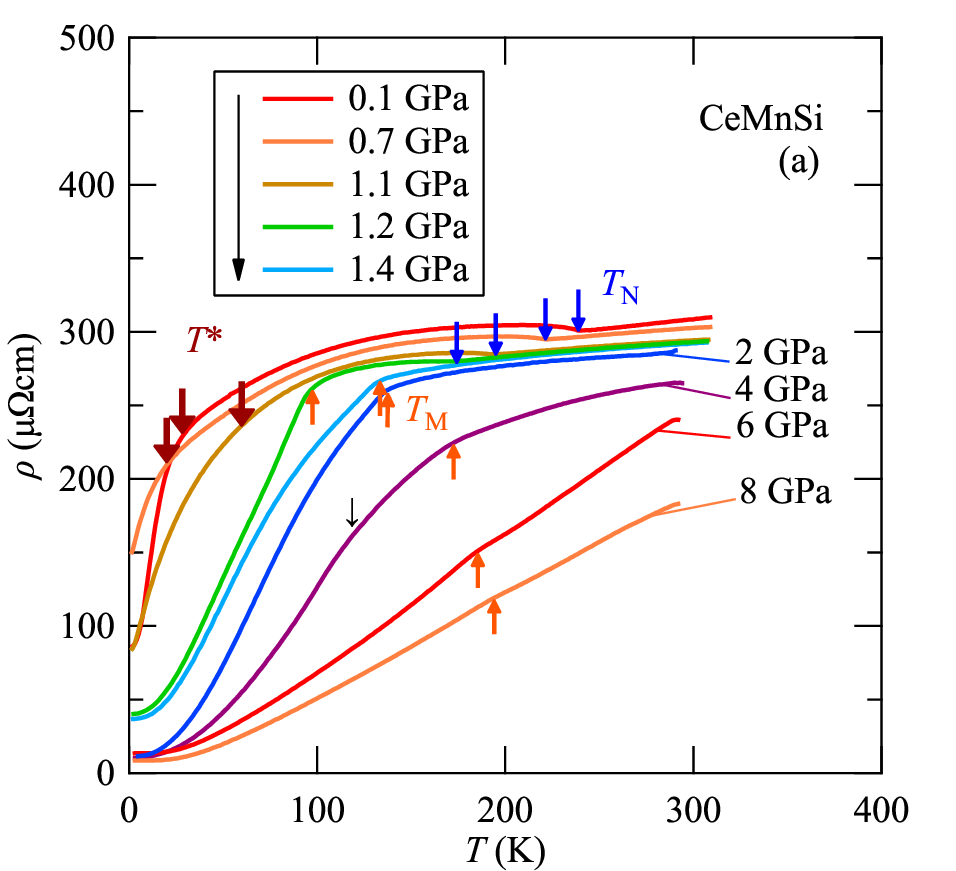}
\includegraphics[width=0.8\linewidth]{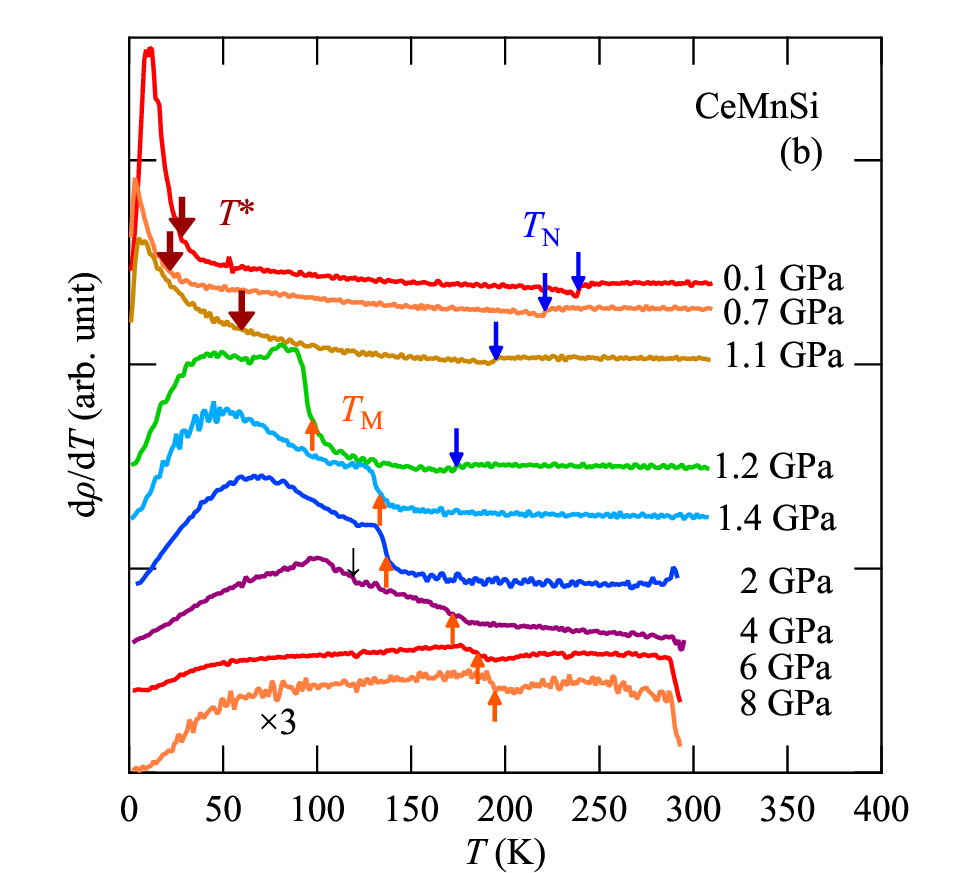}
\caption{(Color online) (a) Temperature dependence of \(\rho\) of CeMnSi under various pressures. (b) Temperature dependence of \(\mathrm{d}\rho/\mathrm{d}T\). 
For clarity, the \(\mathrm{d}\rho/\mathrm{d}T\) curves are vertically offset.
The arrows indicate characteristic temperatures and are guides to the eye (see text).
}
\label{Ele}
\end{center}
\end{figure}

Figure \ref{Ele}(a) shows the temperature dependence of \(\rho\) of CeMnSi under various pressures.
The characteristic temperatures discussed below are determined from anomalies in the temperature derivative
\((\mathrm{d}\rho/\mathrm{d}T)\), as shown in Fig.~\ref{Ele}(b).
 \(T_{\mathrm{N}}\) and \(T_{\mathrm{M}}\) correspond to points of the steepest slope,
while \(T^{\ast}\) is defined as the onset temperature where the slope of
\(\mathrm{d}\rho/\mathrm{d}T\) starts to increase rapidly.

At low pressures, the temperature dependence of \( \rho \) is consistent with previous reports at ambient pressure~\cite{Tanida2023}. At 0.1 GPa, \( \rho \) decreases upon cooling from room temperature down to \( T_{\mathrm{N}} \sim 240 \) K, where a sharp upturn appears. Upon further cooling, \( \rho \) exhibits a broad maximum around 200 K and a shoulder-like feature at \( T^{\ast} \sim 30 \) K.

With increasing pressure, the upturn in \( \rho \) below \( T_{\mathrm{N}} \) is progressively suppressed. 
\( T_{\mathrm{N}} \) decreases monotonically and the associated anomaly in \( \rho \) is no longer discernible above 1.2 GPa, indicating that the antiferromagnetic order disappears above this pressure, which corresponds to the critical pressure \(P_{\mathrm{c}} \sim 1.3\)~GPa.
The shoulder-like feature at \( T^{\ast} \), observed at lower pressures, is also suppressed and disappears at 1.2 GPa. 
In a narrow pressure range between 1.1 and 1.2 GPa, the behavior of \( \rho(T) \) changes abruptly. At 1.2 GPa, a new anomaly appears at \( T_{\mathrm{M}} \sim 97 \) K as a clear kink in \( \rho(T) \), whereas no such anomaly is observed at 1.1 GPa.
These observations indicate that the suppression of the Mn antiferromagnetic order is accompanied by a sudden reorganization of the Mn-derived magnetic state, rather than a continuous evolution.

At intermediate pressures (1.4--4 GPa), the temperature dependence of \( \rho \) deviates markedly from simple metallic behavior. Instead of a monotonic decrease upon cooling, \( \rho(T) \) exhibits a broad, rounded curvature over a wide temperature range, suggesting an additional contribution to the scattering processes. This behavior is absent in the non-\(4f\) reference compound LaMnSi~\cite{Tanida2022}, indicating that the observed feature originates predominantly from the Ce-\(4f\) electrons.

In the higher-pressure region, a further qualitative change in \( \rho(T) \) occurs across the structural transition pressure \( P_{\mathrm{s}} \sim 5.7 \) GPa~\cite{Kawamura2026}. 
The anomaly at \( T_{\mathrm{M}} \) persists up to the highest pressure measured and shifts to higher temperatures with increasing pressure, although it becomes progressively less pronounced at high pressures.
At 6 GPa, \( \rho(T) \) becomes considerably smoother and exhibits a more conventional metallic temperature dependence. The broad curvature observed at lower pressures is strongly suppressed, indicating a substantial modification of the electronic state across \( P_{\mathrm{s}} \).

A weak anomaly is also observed around 4 GPa, appearing as a subtle change in slope in \( \rho(T) \) and a corresponding feature in \( \mathrm{d}\rho/\mathrm{d}T \) near 120 K. Considering the pressure uncertainty and the first-order nature of the structural transition, this anomaly likely reflects the onset of the high-pressure phase at low temperatures.

Taking all these observations together, these results indicate that the temperature dependence of \(\rho\) evolves qualitatively with pressure.
In the low-pressure region, the marked reduction in \(\rho\) below \(T_{\mathrm{M}}\) suggests a significant suppression of scattering, whereas in the high-pressure region, the weak temperature dependence of \(\rho(T)\) is indicative of more conventional metallic behavior.
This evolution implies a qualitative change in the dominant scattering processes with increasing pressure.

\begin{figure}[tbp]
\begin{center}
\includegraphics[width=0.49\linewidth]{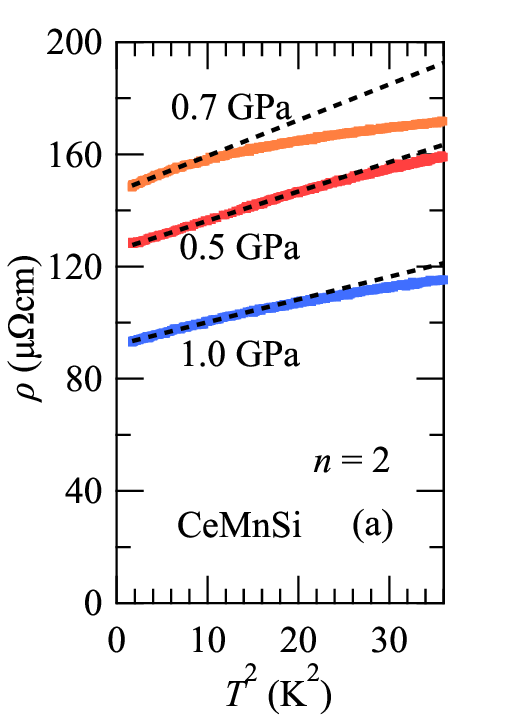}
\includegraphics[width=0.49\linewidth]{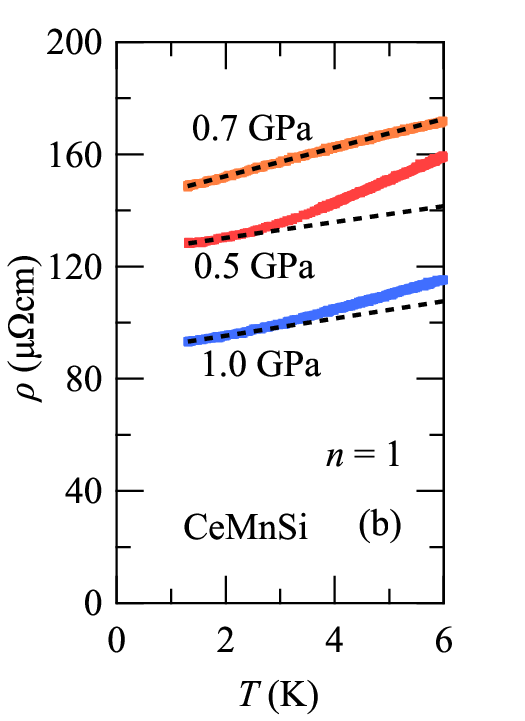}
\includegraphics[width=0.8\linewidth]{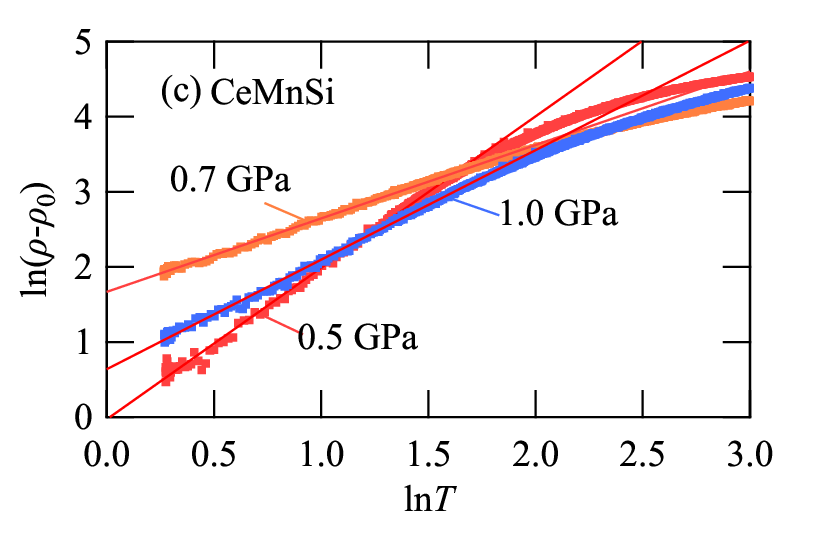}
\includegraphics[width=0.6\linewidth]{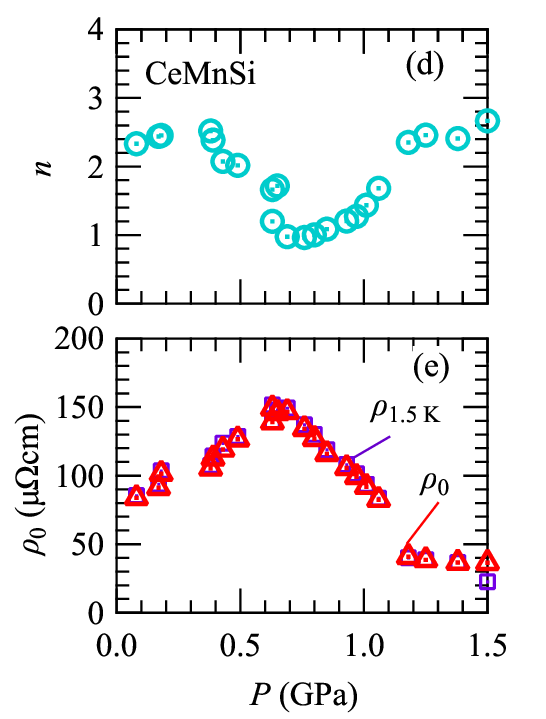}
\caption{(Color online)(a) Low-temperature \(\rho\) of CeMnSi plotted as a function
of \(T^2\) at representative pressures.
(b) \(\rho\) plotted as a function of \(T\) at the same pressures; dashed lines in (a) and (b) are guides to the \(T^2\) and \(T\) dependences, respectively.
(c) Plot of \(\ln(\rho-\rho_0)\) versus \(\ln T\); the solid line is a linear fit according to \(\ln(\rho-\rho_0)=\ln A + n\ln T\).
(d) Pressure dependence of the exponent \(n\).
(e) Pressure dependence of the residual resistivity \(\rho_0\) and resistivity at 1.5 K,  \(\rho_\mathrm{1.5 K}\) of CeMnSi.}
\label{Ele_n}
\end{center}
\end{figure}

Figures~\ref{Ele_n}(a) and \ref{Ele_n}(b) show the low-temperature \(\rho\) of CeMnSi
plotted as functions of \(T^2\) and \(T\), respectively, for representative pressures
of 0.5, 0.7, and 1.0 GPa.
At 0.5 GPa, \(\rho\) follows a linear dependence on \(T^2\) over a wide
temperature range, indicating dominant Fermi-liquid-like transport behavior.
In contrast, at 0.7 GPa, the \(T^2\) dependence is limited to a narrow low-temperature
region, whereas a linear dependence on \(T\) extends over a significantly broader
temperature range.

In order to characterize this evolution in a phenomenological manner, the low-temperature resistivity was analyzed using the empirical power-law form \(\rho=\rho_0+AT^n\).
Here, \(\rho_0\) represents the residual resistivity, and the exponent \(n\) serves as an effective parameter describing deviations from conventional Fermi-liquid behavior (\(n\) = 2)~\cite{Stewart1984,Gegenwart2008}.

Figure~\ref{Ele_n}(c) shows plots of \(\ln(\rho-\rho_{0})\) versus \(\ln T\) for representative
pressures of \(0.5\), \(0.7\), and \(1.0~\mathrm{GPa}\), where the slope directly yields the
resistivity exponent \(n\) according to
\(\ln(\rho-\rho_{0})=\ln A+n\ln T\).
The extracted pressure dependence of \(n\) is summarized in Fig.~\ref{Ele_n}(d).
With increasing pressure, \(n\) decreases from values close to \(n\sim 2\) at lower pressures
and approaches \(n\sim 1\) around \(0.7~\mathrm{GPa}\), indicating a crossover from
conventional Fermi-liquid-like transport to an extended nearly \(T\)-linear resistivity.
Upon further increasing pressure, \(n\) gradually recovers toward larger values.

Figure~\ref{Ele_n}(e) displays the pressure dependence of \(\rho_{0}\), obtained from the phenomenological fitting, together with the resistivity
measured at the lowest experimental temperature, \(\rho_{1.5\,\mathrm{K}}\).
Remarkably, \( \rho_{0} \) and \( \rho_{1.5\,\mathrm{K}} \) not only exhibit the same pressure dependence, including a pronounced maximum around \( 0.7\,\mathrm{GPa} \), but also show nearly identical absolute values over the entire pressure range. This close agreement suggests that the contribution from inelastic scattering at \( 1.5\,\mathrm{K} \) is already very small within the present experimental accuracy, and thus supports the reliability of the extracted \( \rho_{0} \).

Notably, the pressure region where \(\rho_{0}\) is enhanced coincides with that where the
exponent \(n\) approaches \(n\sim 1\), corresponding to the most pronounced deviation from
conventional Fermi-liquid behavior.
It should be emphasized that analyses based on the empirical expression
\(\rho=\rho_{0}+AT^{n}\) are generally more reliable when performed at much lower temperatures,
where the determination of \(\rho_{0}\) becomes less ambiguous.
In the present case, the exponent \(n\) should therefore be regarded as an effective parameter
characterizing deviations from conventional metallic behavior, rather than as a strict
Fermi-liquid exponent defined in a well-established paramagnetic ground state.
Nevertheless, the nearly \(T\)-linear resistivity around \(0.7~\mathrm{GPa}\) is already clearly visible in the \(\rho(T)\) data shown in Fig.~\ref{Ele_n}(b) and does not rely on the fitting procedure.
Despite the pronounced changes in the low-temperature transport properties, no signature of superconductivity was observed down to \( 1.3\,\mathrm{K} \) over the entire pressure range investigated.

Taking all these observations together, the temperature dependence of \( \rho \) evolves systematically with pressure, accompanied by distinct changes in the magnetic state. 
With increasing pressure, the antiferromagnetic ordering at \( T_{\mathrm{N}} \) is progressively suppressed and eventually disappears at \( P_{\mathrm{c}} \). 
In a narrow pressure region near \( P_{\mathrm{c}} \), a new anomaly appears at \( T_{\mathrm{M}} \), indicating a reconstruction of the magnetic state. 
Along with these changes, the temperature dependence of \( \rho \) evolves from behavior consistent with a heavy-fermion state at low pressures, through a region with anomalous temperature dependence around \( P_{\mathrm{c}} \), including a nearly \(T\)-linear behavior. 
At higher pressures, \( \rho(T) \) becomes smoother and exhibits a more conventional metallic behavior, particularly above \(P_{\mathrm{s}}\).

\subsection{Powder XRD at Ambient Pressure} 

\begin{figure*}[tbp]
\begin{center}
\includegraphics[width=0.49\linewidth]{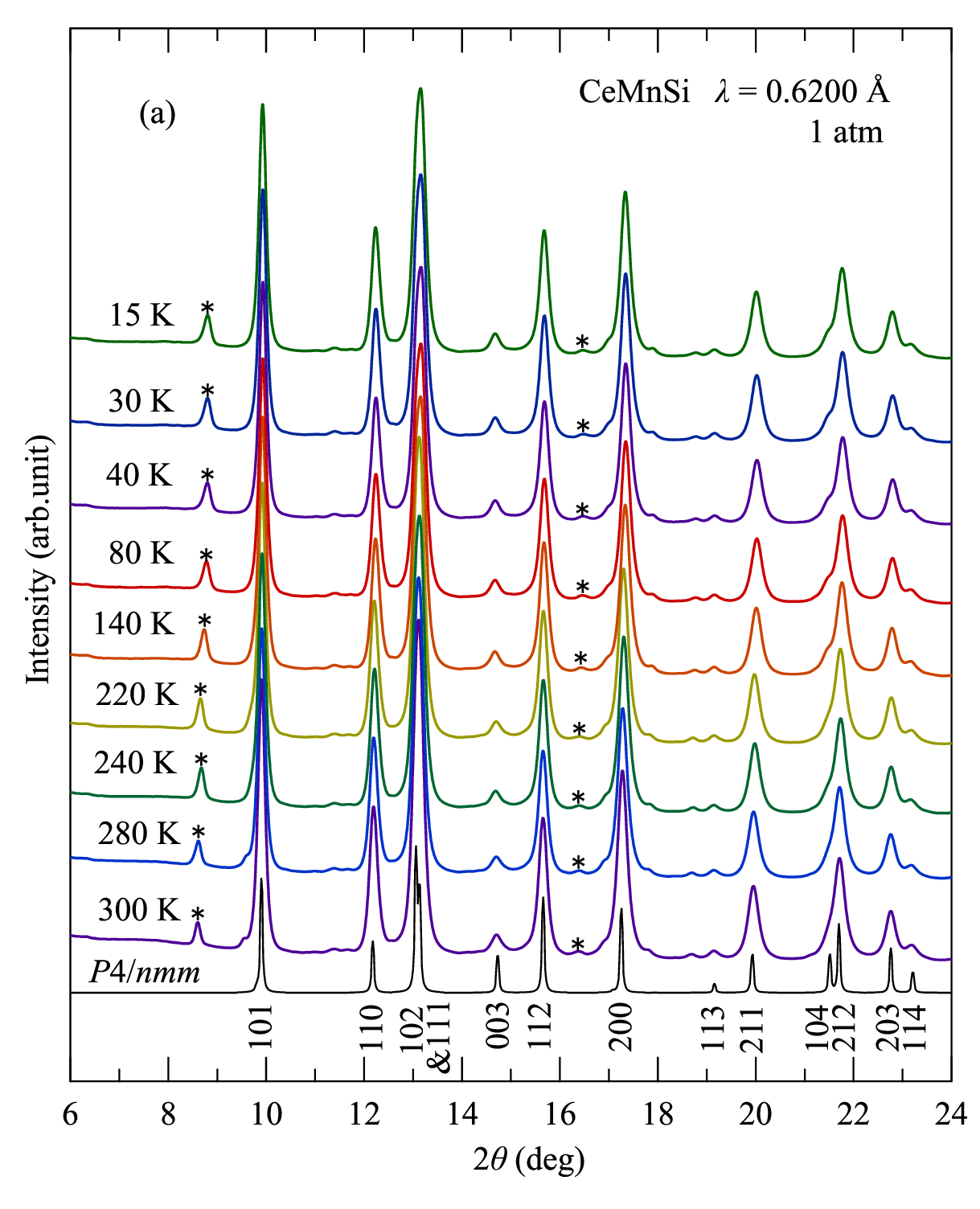}
\includegraphics[width=0.49\linewidth]{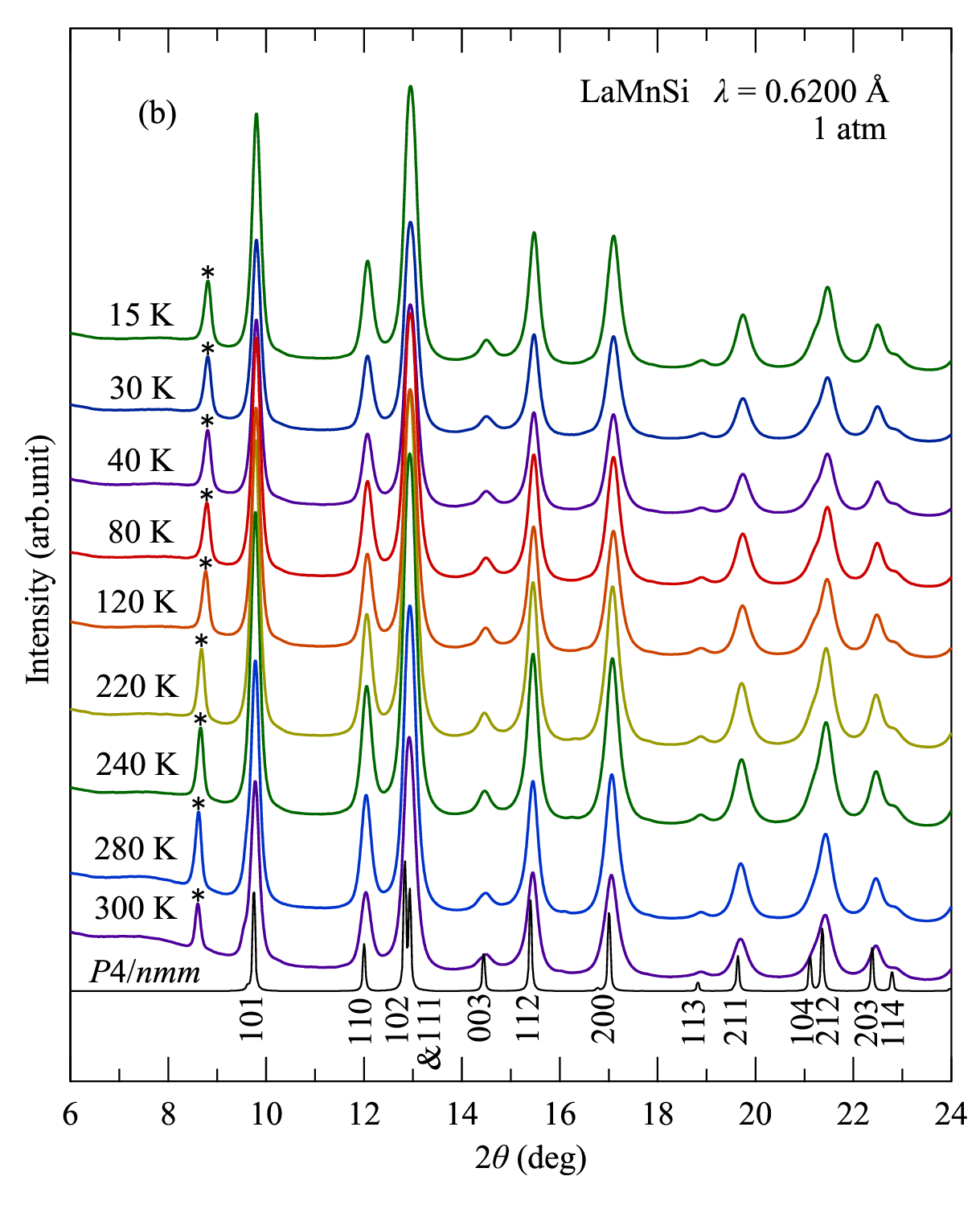}
\includegraphics[width=0.325\linewidth]{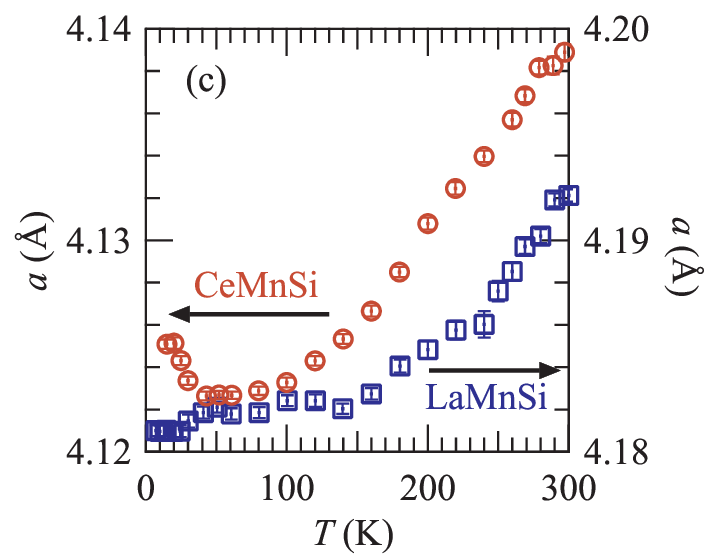}
\includegraphics[width=0.325\linewidth]{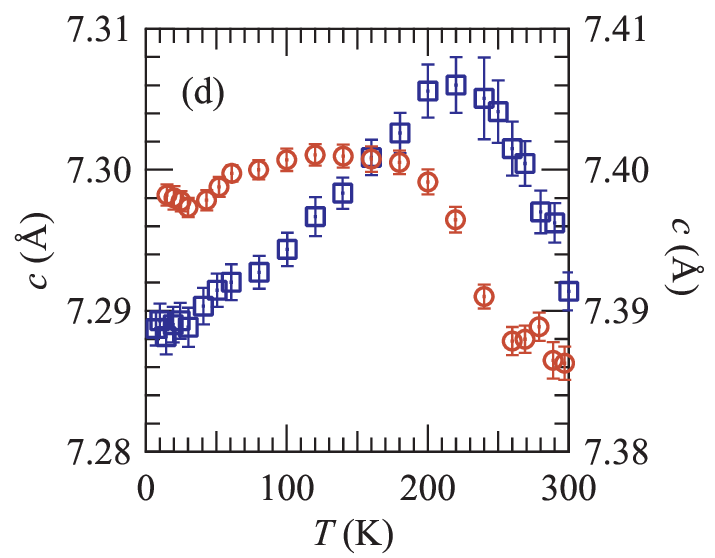}
\includegraphics[width=0.34\linewidth]{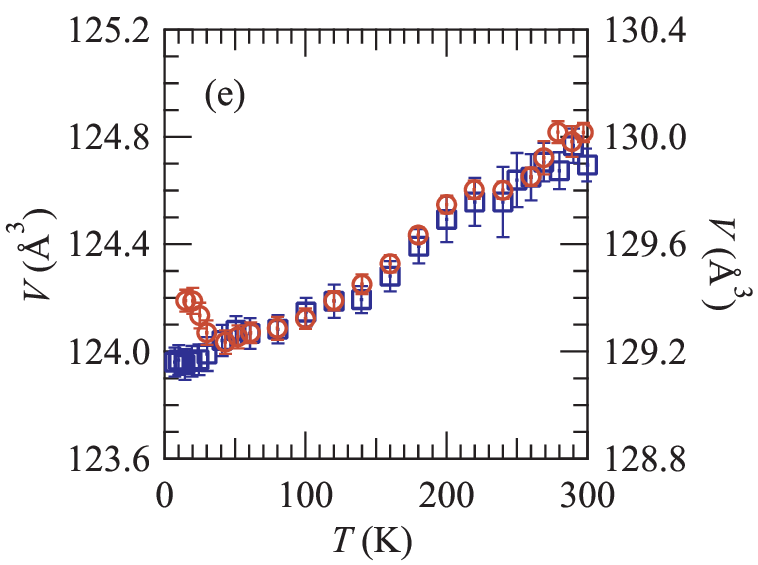}
\caption{(Color online) Temperature-dependent XRD powder patterns of (a) CeMnSi and (b) LaMnSi measured at 1 atm. The calculated diffraction patterns for the \(P4/nmm\) space group are also shown. Asterisks indicate diffraction peaks originating from the Mylar window of the GM refrigerator and Apiezon grease. Temperature dependences of (c) the lattice constant \(a\), (d) the lattice constant \(c\), and (e) the unit-cell volume \(V\) for CeMnSi and LaMnSi at 1 atm.}
\label{XRD_1atm}
\end{center}
\end{figure*}

Figure~\ref{XRD_1atm}(a) shows the temperature-dependent XRD patterns of CeMnSi
measured at ambient pressure.
Asterisks indicate additional diffraction peaks mainly originating from the Apiezon grease, with minor contributions from the Mylar window. 
These peaks are absent in the high-pressure measurements, reflecting the absence of Apiezon grease.
All diffraction peaks can be indexed by the tetragonal $P4/nmm$ space group,
and no peak splitting or emergence of extra reflections is observed over the
entire temperature range from 15 to 300~K. A comparison of the diffraction patterns at 15~K and 300~K reveals that no pronounced redistribution of Bragg peak intensities is observed within experimental accuracy.
This indicates that no major modification of the crystallographic
structure, such as a symmetry lowering or a large atomic displacement,
occurs at ambient pressure.
Similar behavior is observed for LaMnSi, as shown in
Fig.~\ref{XRD_1atm}(b).

The temperature dependence of the lattice parameters of CeMnSi is summarized in
Figs.~\ref{XRD_1atm}(c) and \ref{XRD_1atm}(d).
With decreasing temperature, the lattice constant $a$ decreases monotonically
from 300~K down to approximately 100~K, remains nearly temperature independent
between 80 and 40~K, and exhibits a slight but discernible increase below 40~K.
In contrast, the lattice constant $c$ shows a gradual increase upon cooling
from 300~K to 200~K, followed by an almost constant value down to 60~K,
a weak decrease down to 30~K, and a subsequent increase toward the lowest
measured temperature.
Note that, in ordinary compounds,
both lattice parameters and the unit-cell volume are expected
to decrease monotonically upon cooling; the simultaneous increase of both $a$ and $c$ at low temperatures is highly anomalous.

Figure~\ref{XRD_1atm}(e) presents the temperature dependence of the unit-cell
volume $V$.
In CeMnSi, the concurrent increase of both $a$ and $c$ below approximately 40~K results in a small but discernible negative thermal expansion at low temperatures.
From the temperature variation of the lattice parameters,
the magnitude of the negative thermal expansion, expressed as the relative volume change \(\Delta V/V\), is roughly estimated
to be of the order of \(2 \times 10^{-3}\).
Although this estimate is based on a simple comparison between 40~K
and the lowest measured temperature,
it provides a useful measure of the anomalous lattice response in CeMnSi.

For LaMnSi, the lattice constant $a$ decreases almost monotonically over the
entire temperature range, as shown in Fig.~\ref{XRD_1atm}(c).
The lattice constant $c$ increases from 300~K to about 220~K and then decreases smoothly upon further cooling, as displayed in Fig.~\ref{XRD_1atm}(d).
Although \(c\) exhibits a nonmonotonic temperature dependence, the unit-cell volume decreases monotonically with decreasing temperature over the entire temperature range. 
Accordingly, no negative thermal expansion is observed in LaMnSi, in clear contrast to CeMnSi, as shown in Fig.~\ref{XRD_1atm}(e).

Similar anisotropic temperature dependences of the lattice parameters
have been reported in the isostructural compound CeCoSi%
~\cite{Tanida2019,Matsumura2022},
where the lattice constant $a$ decreases smoothly whereas $c$ increases
upon cooling down to the hidden-order temperature \(T_\mathrm{0}\sim12\) K~\cite{Matsumura2022}.
The broad maximum found in Fig.~\ref{XRD_1atm}(d) may thus be a common
feature within the $R T$Si family.

Although the dip in the lattice parameter \(c\) around 30~K
appears relatively sharp,
it does not correspond to a phase transition; no anomaly has been detected in bulk properties such as \(\rho\) thus far.
Therefore, the marked negative thermal expansion observed below 30~K should be regarded as an intrinsic lattice response of CeMnSi rather than a structural instability.
Given that this anomalous volumetric expansion is absent in the non-\(4f\) reference compound LaMnSi, it is most likely related to the development of a nontrivial heavy-fermion state associated with the Ce-\(4f\) electrons.

\subsection{Powder XRD under Pressure} 

In our recent high-pressure XRD study, CeMnSi was found to undergo a first-order structural transition from the tetragonal \(P4/nmm\) phase to the monoclinic \(P2_1/m\) phase at \(P_{\mathrm{s}} \sim 5.7~\mathrm{GPa}\) at room temperature~\cite{Kawamura2026}.
On the other hand, the present study has revealed unique low-temperature transport properties at low pressures and a marked negative thermal expansion at ambient pressure.
In the present study, motivated by the low-temperature transport anomalies and negative thermal expansion revealed above, we extend XRD measurements to low temperatures in order to establish the temperature--pressure phase diagram.

\begin{figure}
\begin{center}
\includegraphics[width=\linewidth]{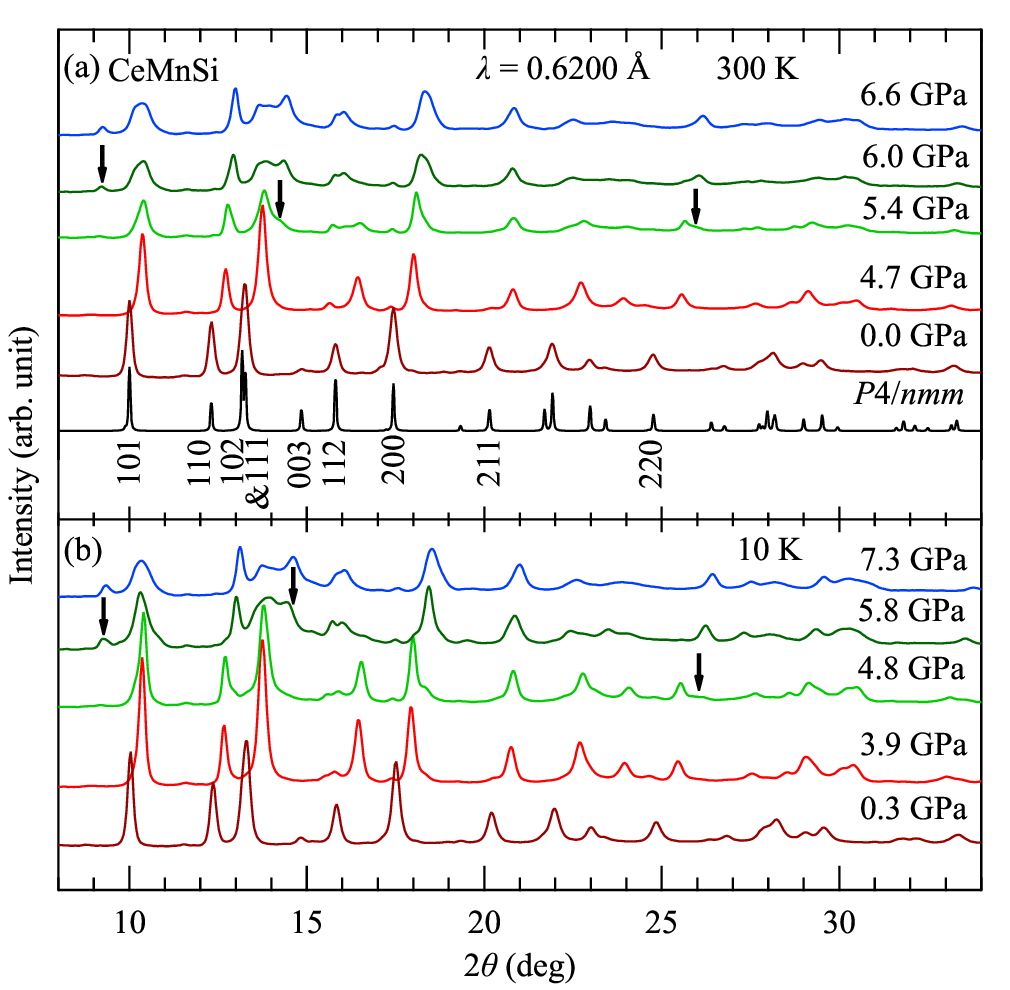}
\caption{(Color online) XRD patterns of CeMnSi under pressure measured at (a) 300~K and (b) 10~K. Arrows indicate characteristic diffraction peaks of the high-pressure phase, marked near their respective onset pressures. The simulated diffraction pattern for the tetragonal \(P4/nmm\) space group is also shown for comparison.}
\label{XRD}
\end{center}
\end{figure}

Figures~\ref{XRD}(a) and \ref{XRD}(b) show the XRD patterns measured at 300~K and 10~K, respectively.
At both temperatures, diffraction patterns at low pressures are well
described by the tetragonal $P4/nmm$ structure.
With increasing pressure, additional diffraction features develop, and clear changes in the diffraction profiles are observed above approximately \(5.4\)~GPa at 300~K and \(4.8\)~GPa at 10~K. 
In particular, the appearance of the 100 reflection, forbidden in the \(P4/nmm\) symmetry, provides direct evidence for the emergence of a symmetry-lowered phase.

\begin{figure}
\begin{center}
\includegraphics[width=0.8\linewidth]{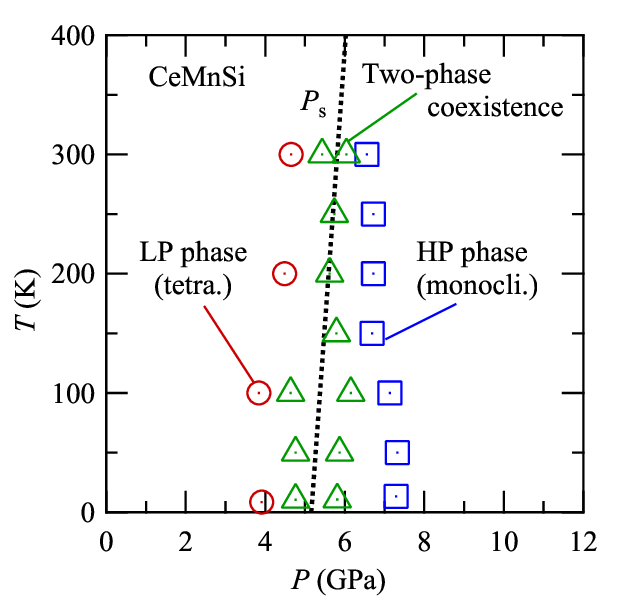}
\caption{(Color online) Temperature--pressure phase diagram of CeMnSi constructed from the powder XRD measurements.
The low-pressure tetragonal phase (circles) and the high-pressure monoclinic phase (squares) are separated by a first-order structural transition, accompanied by a two-phase coexistence region (triangles) near \(P_{\mathrm{s}}\).
The dashed line is a guide to the eye.}
\label{XRDPT}
\end{center}
\end{figure}

At 10~K, the diffraction patterns indicate that the tetragonal and monoclinic phases are stabilized in distinct pressure regions, while both phases coexist in an intermediate pressure range, consistent with a first-order structural transition. 
The structural transition pressure \(P_{\mathrm{s}}\) is defined as the midpoint of this coexistence region.
At 300~K, \(P_{\mathrm{s}}\) is estimated to be \(5.7 \pm 0.3\)~GPa, consistent with previous work~\cite{Kawamura2026}, whereas it shifts slightly to lower pressures, \(5.3 \pm 0.5\)~GPa, at 10~K. 
The resulting temperature--pressure phase diagram is summarized in Fig.~\ref{XRDPT}.

\subsection{Temperature--pressure phase diagram}  

\begin{figure}[t]
\begin{center}
\includegraphics[width=\linewidth]{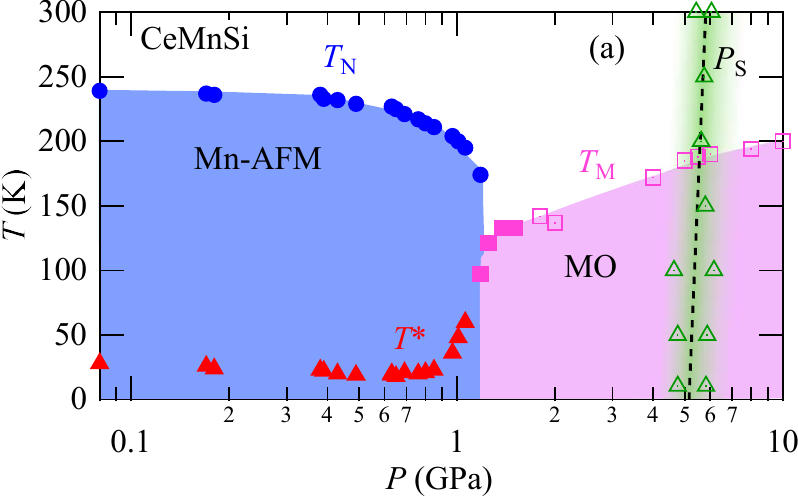}
\caption{(Color online) Temperature--pressure phase diagram of CeMnSi. The pressure axis is plotted on a logarithmic scale to display low-pressure features
around the suppression of \(T_{\mathrm N}\). The characteristic temperatures \(T_{\mathrm N}\), \(T^{\ast}\), and \(T_{\mathrm M}\) were determined
from electrical-resistivity measurements, while the structural transition pressure \(P_{\mathrm s}\) was obtained from the XRD measurements shown in Fig.~\ref{XRDPT}. 
Filled symbols denote data obtained using a piston--cylinder cell, whereas open symbols represent measurements performed with a cubic-anvil apparatus. Here, \(T_\mathrm{N}\) is the antiferromagnetic transition temperature, \(T^{\ast}\) corresponds to the onset temperature of an anomaly in \(\mathrm{d}\rho/\mathrm{d}T\), and  \(T_\mathrm{M}\) denotes an additional anomaly of unclear origin. The lines are guides to the eye.}
\label{Ele_PT}
\end{center}
\end{figure}

Figure~\ref{Ele_PT} shows the temperature--pressure (\(T\)--\(P\)) phase diagram of CeMnSi determined by \(\rho\) and XRD measurements. With increasing pressure,  \(T_{\mathrm N}\) decreases gradually at low pressures, and is rapidly suppressed above 0.7 GPa, then disappears at \(P_{\mathrm c} \sim 1.3\)~GPa. At the same time, \(T_{\mathrm M}\) suddenly appears at 1.2 GPa. 
In a narrow pressure region just below \(P_{\mathrm c}\), we clearly observed two distinct anomalies in \( \rho \), corresponding to \(T_{\mathrm N}\) and \(T_{\mathrm M}\), which are identified in \(\mathrm{d}\rho/\mathrm{d}T\) (Fig.~\ref{Ele}(b)).
The low-temperature \( \rho \) exhibits non-Fermi-liquid-like behavior near \(P = 0.7\) GPa. 
This behavior precedes the significant changes in the Mn-antiferromagnetic ordering, indicating the presence of marked fluctuations in the low-temperature electronic state.

Note that the onset pressure where  \(T_{\mathrm N}\) begins to be suppressed and the pressure where the non-Fermi-liquid-like behavior becomes most pronounced appear to coincide with each other. The pressure dependence of \(T^{\ast}\) may reflect these drastic changes in the electronic states; \(T^{\ast}\) is nearly \(P\)-independent until 0.4 GPa, followed by a minimum at 0.7 GPa, above which \(T^{\ast}\) increases as pressure approaches \(P_{\mathrm c} = 1.3\)~GPa.

Above 1.8 GPa, \(T^{\ast}\) disappears, and \(T_{\mathrm M}\) increases gradually from 140 K to 200 K with increasing pressure up to at least 10 GPa, regardless of the structural transition at \(P_{\mathrm s}\). Although the nature of the phase below \(T_{\mathrm M}\) is not yet clarified, a distinct phase region (denoted as MO in Fig.~\ref{Ele_PT}) is identified from the \(\rho\) anomaly at \(T_{\mathrm M}\). The pressure dependence of \(T_{\mathrm M}\) suggests that the underlying order parameter is different from that at \(P_{\mathrm s}\). A similar high-pressure phase has also been reported in related LaMnSi~\cite{Aoyama}.

In addition, above 5 GPa, the structural transition occurs at \(P_{\mathrm s}\). 
The transition pressure \(P_{\mathrm s}\) shows a weak temperature dependence, taking \(5.7 \pm 0.3\) GPa at room temperature and \(5.3 \pm 0.5\) GPa at low temperatures. 
Notably, this structural transition is observed only in \(R\)=Ce among the \(R\)MnSi series, suggesting a possible relationship with the presence of the Ce-\(4f\) electrons.

\section{Discussion}

The results demonstrate that pressure induces a reorganization of magnetic order, electronic correlations, and the lattice in CeMnSi, leading to a rich temperature--pressure phase diagram.
This behavior reflects the difference in energy scales between the high-temperature Mn-derived magnetism and the low-temperature Ce-\(4f\) heavy-fermion state, indicating that the Mn magnetic order provides the electronic environment for the Ce-\(4f\) electrons under pressure.

\subsection{Anisotropic thermal expansion and negative thermal expansion}

In the high-temperature region, the lattice parameters of \(R\)MnSi (\(R =\) La, Ce) exhibit anisotropic temperature dependences, as shown in Fig.~\ref{XRD_1atm}. 
The lattice parameter \(a\) decreases monotonically with decreasing temperature in both compounds, as expected for ordinary thermal contraction. 
In contrast, the lattice parameter \(c\) exhibits a more complex temperature dependence, increasing upon cooling at high temperatures in both compounds and subsequently decreasing in LaMnSi, whereas it remains nearly temperature-independent over a wide temperature range in CeMnSi.

Similar anisotropic thermal expansion has been reported in the related compound CeCoSi, where \(c\) increases while \(a\) decreases upon cooling down to the hidden-order transition temperature \(T_0\)~\cite{Matsumura2022,Kawamura2022}. 
These observations indicate that such anisotropic lattice responses are a common feature of the \(RT\)Si family. 
As a result of this anisotropy, the unit-cell volume \(V\) decreases smoothly with decreasing temperature at high temperatures, as shown in Fig.~\ref{XRD_1atm}(e), indicating that no anomalous volumetric effect is present in this temperature range.

In contrast to the high-temperature behavior, a qualitatively different lattice response emerges at low temperatures in CeMnSi. 
Below approximately \(40\)~K, both lattice parameters \(a\) and \(c\) exhibit a slight increase upon further cooling, resulting in a clear increase in the unit-cell volume \(V\). 
This behavior corresponds to a volumetric negative thermal expansion observed only in CeMnSi, whereas no such behavior is found in LaMnSi. 
No peak splitting or significant broadening is observed in the Bragg reflections in this temperature range, excluding a structural phase transition as its origin. 
Furthermore, no clear anomaly indicative of a phase transition has been reported in bulk properties such as \(\rho\) and magnetic susceptibility~\cite{Tanida2023}. 
These results demonstrate that the negative thermal expansion is an intrinsic property of CeMnSi rather than a consequence of a structural instability.

In Ce-based heavy-fermion compounds, negative thermal expansion is often observed at low temperatures and is generally associated with the strong volume dependence of the characteristic electronic energy scale, such as the Kondo temperature~\cite{Fetisov1985, Hackl2008, deVisser1989, Takeuchi2014}. 
In CeMnSi, the onset of the negative thermal expansion below \(\sim 40\)~K coincides with characteristic features in the electronic properties, including a steep decrease in \(\rho\) and a broad shoulder or maximum in magnetic susceptibility~\cite{Tanida2023}, which are commonly regarded as signatures of heavy-fermion formation. 
These observations suggest that the negative thermal expansion in CeMnSi is closely related to the development of a nontrivial heavy-fermion state associated with the Ce-\(4f\) electrons.

It is also noteworthy that a field-induced ordered phase (FIOP) appears below \(\sim 30\)~K under magnetic fields applied parallel to the \(a\) axis~\cite{CMS2024}. 
Although the relationship between the FIOP and the negative thermal expansion is not yet clarified, both phenomena may reflect a strong coupling between the electronic state and the lattice in CeMnSi. 
Further experimental and theoretical studies are required to elucidate the interplay between these phenomena.

\subsection{Non-Fermi-liquid behavior below \(P_{\mathrm{c}}\)}

The temperature dependence of \(\rho\) in CeMnSi at ambient pressure is ascribed to Kondo scattering at high temperatures and the formation of a heavy-fermion state at low temperatures~\cite{Tanida2023}. 
Below \(P_{\mathrm{c}} \sim 1.3\) GPa, where the Mn antiferromagnetic order persists, \(\rho(T)\) remains similar to that at ambient pressure. 
The characteristic temperature \(T^{\ast}\), associated with heavy-fermion formation, decreases gradually with increasing pressure, although the Kondo energy scale is generally expected to increase under pressure in Ce-based compounds. 

On the other hand, in the vicinity of \(0.7\) GPa, non-Fermi-liquid behavior characterized by nearly \(T\)-linear resistivity appears, accompanied by an enhancement of the \(\rho_0\). 
Despite some ambiguity in determining \(T^{\ast}\), its decrease may be related to the emergence of NFL behavior at low temperatures. 
In conventional Doniach-type systems, non-Fermi-liquid behavior is often associated with the suppression of magnetic order~\cite{Stewart1984,Lohneysen2007}.
More generally, non-Fermi-liquid behavior can arise from magnetic instabilities and the associated critical fluctuations in correlated electron systems~\cite{Varma2016}.
In CeMnSi, however, Ce-4\(f\) electrons do not form long-range magnetic order even at ambient pressure~\cite{Tanida2023}, whereas Mn-3\(d\) electrons exhibit antiferromagnetic ordering. 

With increasing pressure, the \(T_{\mathrm{N}}\) begins to be suppressed rapidly above \(0.7\) GPa and disappears at \(P_{\mathrm{c}} \sim 1.3\) GPa, where a new anomaly at \(T_{\mathrm{M}}\) emerges. 
These observations suggest the presence of magnetic instability in the low-pressure region below \(P_{\mathrm{c}}\). 
Although the energy scale of the Mn ordering is as high as \(100\) K, low-energy fluctuations may be induced in the Ce-\(4f\) electronic state via Mn-\(3d\)--Ce-\(4f\) hybridization. 
Therefore, the non-Fermi-liquid behavior observed around \(0.7\) GPa can be regarded as a manifestation of this magnetic instability, likely mediated by low-energy fluctuations associated with the destabilization of the Mn antiferromagnetic order.

It is noteworthy that, as seen in \(\rho(T)\) at \(P = 1.2\) GPa, the temperature dependence remains similar to that at ambient pressure above \(T_{\mathrm{M}}\), whereas \(\rho\) shows a marked suppression below \(T_{\mathrm{M}}\). 
This indicates that the emergence of \(T_{\mathrm{M}}\) significantly affects the electronic state.

The microscopic origin of the nearly \(T\)-linear resistivity observed around \(0.7~\mathrm{GPa}\) remains unclear. 
However, this pressure region coincides with the onset of the suppression of the antiferromagnetic order of the Mn sublattice, suggesting that the anomalous transport behavior is closely related to the magnetic instability. 
From this viewpoint, low-energy fluctuations associated with the destabilization of the Mn magnetic order may play an important role in the emergence of the non-Fermi-liquid behavior.

This correspondence is particularly noteworthy because magnetic order is directly linked to the symmetry of the electronic state. 
Therefore, the pressure-induced change in the Mn magnetic state may involve not only a modification of the magnetic ordering but also a change in the underlying symmetry of the system, possibly including the breaking of the \(PT\) symmetry suggested at ambient pressure. 
Consequently, the emergence of the anomaly at \(T_{\mathrm{M}}\) above \(P_{\mathrm{c}}\) can be understood as a manifestation of a reconstructed magnetic state with a different symmetry.

\subsection{Pressure-induced transition at \(T_{\mathrm{M}}\) and its relation to the structural transition at \(P_{\mathrm{s}}\)}

The transport results presented above indicate that the temperature dependence of \(\rho\) changes qualitatively with pressure, suggesting a modification of the dominant scattering processes. In particular, the pronounced decrease in \(\rho\) below \(T_{\mathrm{M}}\) at low pressures reflects a significant change in the electronic state.
This behavior may be attributed to a suppression of Kondo-type scattering, corresponding to a reconstruction of the Ce-\(4f\) electronic state in the Mn-antiferromagnetic background. This change may be closely related to a modification of the Mn magnetic structure, possibly involving the breaking of the combined \(PT\) symmetry, which stabilizes the heavy-fermion state.

The transport data also give further insight into the relationship between the electronic states and the lattice. At \(P = 4\)~GPa, a subtle change in the slope of \(\rho(T)\) is observed around \(120\)~K, accompanied by a corresponding feature in \(\mathrm{d}\rho/\mathrm{d}T\). Although this anomaly is broad, it may be associated with a precursor of the structural transition at \(P_{\mathrm{s}}\).
This interpretation is consistent with the XRD results, which reveal a two-phase coexistence region near \(P_{\mathrm{s}}\). Considering the pressure uncertainty of approximately \(0.5\)~GPa, the anomaly at 4~GPa likely reflects the onset of the high-pressure monoclinic phase at low temperatures.
Accordingly, the evolution of \(\rho(T)\) at intermediate pressures can be understood as resulting from the competition between the tetragonal and monoclinic phases, accompanied by a reduction of the Ce-\(4f\)-derived scattering.

On this basis, the pressure evolution of the anomaly at \(T_{\mathrm{M}}\) is discussed as follows.
Above \(P \sim 1.2\)~GPa, where the Mn antiferromagnetic order is suppressed, a new anomaly appears at \(T_{\mathrm{M}}\) and shifts to higher temperatures with increasing pressure. Since the energy scale of \(T_{\mathrm{M}}\) is comparable to that of \(T_{\mathrm{N}}\), this anomaly is naturally attributed to a pressure-induced reorganization of the Mn-\(3d\) magnetic state rather than to a low-energy instability of the Ce-\(4f\) electrons.

In the \(R\)MnSi family, the magnetic structure of the Mn sublattice is known to be sensitive to the Mn--Mn interatomic distance, exhibiting distinct antiferromagnetic and ferromagnetic states depending on the rare-earth element~\cite{Welter1994}. 
Recent microscopic studies using high-quality single crystals, such as NMR measurements on LaMnSi~\cite{Sakai2026}, have revealed a C-type antiferromagnetic structure with Mn moments aligned along the tetragonal \(c\) axis. These results highlight the robustness of the layered Mn-derived magnetism and its sensitivity to subtle changes in the lattice parameters. In this context, pressure-induced lattice contraction can modify the Mn--Mn exchange interactions, leading to a reconstruction of the magnetic state, as also suggested by electronic-structure calculations on the \( R \)MnSi family~\cite{Chevalier2004}

The switching of the Mn magnetic structure is expected to modify the symmetry of the system. At ambient pressure, CeMnSi is suggested to realize a \(PT\)-symmetric antiferromagnetic state below \(T_{\mathrm{N}}\), in which spin degeneracy is protected and a nontrivial heavy-fermion state can be formed at low temperatures~\cite{Tanida2023}. If the Mn magnetic structure is altered under pressure, the \(PT\) symmetry may be lifted, resulting in a reconstruction of the electronic band structure and a destabilization of the heavy-fermion state. From this point of view, the marked suppression of \(\rho\) below \(T_{\mathrm{M}}\) can be interpreted as a strong reduction in the scattering between conduction electrons and the Ce-\(4f\) electrons, reflecting a strong suppression of the heavy-fermion state. The relatively sharp drop in \(\rho\) may favor a scenario involving enhanced ferromagnetic correlations at \(T_{\mathrm{M}}\), although a definitive identification of the magnetic structure requires microscopic probes. In this context, it is noteworthy that a pressure-induced evolution toward ferromagnetic order has recently been reported in the related compound LaMnSi~\cite{Aoyama}, supporting the idea that the Mn--Mn exchange interactions in this material family are highly sensitive to pressure.

At higher pressures, \(T_{\mathrm{M}}\) continues to increase even across the structural transition at \(P_{\mathrm{s}}\), indicating that the underlying magnetic instability of the Mn sublattice persists over a wide pressure range. On the other hand, the characteristic \(\rho(T)\) drop below \(T_{\mathrm{M}}\) becomes progressively less pronounced above \(\sim 2\)~GPa and evolves into only a weak kink-like anomaly above \(P_{\mathrm{s}}\). In addition, the broad shoulder in \(\rho(T)\) observed around \(150\)~K in the low-pressure region is strongly suppressed above \(P_{\mathrm{s}}\). These results indicate that the electronic state changes qualitatively across the structural transition.

Below \(P_{\mathrm{s}}\), the evolution of \(\rho(T)\) can be primarily understood in terms of a gradual suppression of scattering from the Ce-\(4f\) electrons. Indeed, the temperature dependence of \(\rho\) at intermediate pressures (e.g., \(4\)~GPa) becomes similar to that of LaMnSi at ambient pressure, suggesting that the Ce-\(4f\) electrons become weakly involved in magnetic scattering. Above \(P_{\mathrm{s}}\), \(\rho(T)\) exhibits a more conventional metallic behavior over a wide temperature range. This change indicates a substantial modification of the Fermi surface across the structural transition, which likely suppresses not only the Ce-\(4f\)-derived scattering but also the magnetic scattering associated with the Mn-\(3d\) electrons. Consistently, the pressure dependence of \(T_{\mathrm{M}}\) appears to be slightly weakened above \(P_{\mathrm{s}}\), implying a modification of the underlying magnetic interactions.

It is noteworthy that the structural transition at \(P_{\mathrm{s}}\) is observed only in CeMnSi among the \(R\)MnSi series, highlighting the essential role of the Ce-\(4f\) electrons. 
In analogy with related systems such as CeCoSi, a valence instability of the Ce ions has been proposed as a possible driving mechanism~\cite{Kawamura2022}, and a recent theoretical study based on DFT+DMFT has further pointed out that the pressure-induced structural transition in CeCoSi can be driven by a valence instability of the Ce-\(4f\) electrons~\cite{Zhang2025}. 
However, at least in the case of CeMnSi, Ce-4\( f \) electrons are inferred to play a reduced role in magnetism even at \( P = 4 \) GPa below \( P_{\mathrm{s}} \), as mentioned above. 
This observation suggests that lattice-related effects play an important role in the transition at \( P_{\mathrm{s}} \). 
In this context, a lattice instability associated with the presence of the Ce-4\( f \) electrons may eventually drive the structural transition at \( P_{\mathrm{s}} \).

Taken together, the results of this study indicate that the physical properties of CeMnSi cannot be understood in terms of a single driving mechanism. 
Instead, the negative thermal expansion, non-Fermi-liquid-like transport, pressure-induced anomaly at \(T_{\mathrm M}\), and structural transition consistently point to a strong interplay among Ce-4$f$ electrons, Mn-3$d$ magnetism, and the lattice. 
This interplay provides a unified picture of the complex phase evolution in CeMnSi under pressure.

\section{Summary}

We investigated \(\rho\) and powder XRD in CeMnSi under high pressure in order to clarify the pressure evolution of the heavy-fermion state and its coupling to magnetism and the lattice.  
With increasing pressure, the antiferromagnetic order of Mn-\(3d\) electrons is rapidly suppressed and disappears at \( P_{\mathrm{c}} \sim 1.3 \) GPa. Around \( 0.7 \) GPa, non-Fermi-liquid behavior characterized by nearly \( T \)-linear resistivity develops, indicating enhanced electronic fluctuations.  
Above \( P_{\mathrm{c}} \), an additional anomaly emerges at \( T_{\mathrm{M}} \), which likely originates from a pressure-induced modification of the Mn magnetic ordering. Such a change in the magnetic structure may modify the symmetry of the system and destabilize the heavy-fermion state, possibly through a breaking of the \( PT \) symmetry.  
The temperature \( T_{\mathrm{M}} \) increases monotonically with pressure and persists across the structural transition at \( P_{\mathrm{s}} \sim 5.7 \) GPa.  
At ambient pressure, CeMnSi exhibits pronounced negative thermal expansion below \( \sim 40 \) K, further supporting the realization of a heavy-fermion ground state.

\section*{Acknowledgment}
\begin{acknowledgment}
Synchrotron XRD experiments were performed at BL-18C of the Photon Factory, KEK, with the approval of the Photon Factory Program Advisory Committee (Proposal No. 2023G532, 2025G525). This work was carried out using the facilities of the Institute for Solid State Physics, the University of Tokyo (No. 202304-GNBXX-0067). A part of this work was supported by JSPS KAKENHI Grant Nos. JP22K19076, JP25K01485, and JP25K07209.
\end{acknowledgment}


\begin{thebibliography}{99}

\bibitem{Stewart1984}
G.~R.~Stewart,
Rev.\ Mod.\ Phys.\ \textbf{56}, 755 (1984).

\bibitem{Gegenwart2008}
P.~Gegenwart, Q.~Si, and F.~Steglich,
Nat.\ Phys.\ \textbf{4}, 186 (2008).

\bibitem{Lohneysen2007}
H.~v.~L{\"o}hneysen, A.~Rosch, M.~Vojta, and P.~W{\"o}lfle,
Rev.\ Mod.\ Phys.\ \textbf{79}, 1015 (2007).

\bibitem{Drotziger2006}
S.~Drotziger, C.~Pfleiderer, M.~Uhlarz, H.~v.~L{\"o}hneysen,
D.~Souptel, W.~L{\"o}ser, and G.~Behr,
Phys.\ Rev.\ B \textbf{73}, 214413 (2006).

\bibitem{Luo2015}
Q.~Luo, G.~Garbarino, B.~Sun, D.~Fan, Y.~Zhang, Z.~Wang,
Y.~Sun, J.~Jiao, X.~Li, P.~Li, N.~Mattern, J.~Eckert,
and J.~Shen,
Nat.\ Commun.\ \textbf{6}, 5703 (2015).

\bibitem{Bodak1970}
O.~I.~Bodak, E.~I.~Gladyshevskii, and P.~I.~Kripyakevich,
Zh.\ Strukt.\ Khim.\ \textbf{11}, 283 (1970).

\bibitem{vesta}
K.~Momma and F.~Izumi,
J.\ Appl.\ Crystallogr.\ \textbf{44}, 1272 (2011).

\bibitem{Tanida2019}
H.~Tanida, K.~Mitsumoto, Y.~Muro, T.~Fukuhara, Y.~Kawamura,
A.~Kondo, K.~Kindo, Y.~Matsumoto, T.~Namiki, T.~Kuwai,
and T.~Matsumura,
J.\ Phys.\ Soc.\ Jpn.\ \textbf{88}, 054716 (2019).

\bibitem{Matsumura2022}
T.~Matsumura, S.~Kishida, M.~Tsukagoshi, Y.~Kawamura,
H.~Nakao, and H.~Tanida,
J.\ Phys.\ Soc.\ Jpn.\ \textbf{91}, 064704 (2022).

\bibitem{Tanida2023}
H.~Tanida, H.~Matsuoka, Y.~Kawamura, and K.~Mitsumoto,
J.\ Phys.\ Soc.\ Jpn.\ \textbf{92}, 044703 (2023).

\bibitem{Kawamura2020}
Y.~Kawamura, H.~Tanida, R.~Ueda, J.~Hayashi, K.~Takeda,
and C.~Sekine,
J.\ Phys.\ Soc.\ Jpn.\ \textbf{89}, 054702 (2020).

\bibitem{Kawamura2026}
Y.~Kawamura, S.~Nishiyama, J.~Hayashi, K.~Takeda,
C.~Sekine, and H.~Tanida,
J.\ Phys.\ Soc.\ Jpn.\ \textbf{95}, 014601 (2026).

\bibitem{Pb}
T.~F.~Smith and C.~W.~Chu,
Phys.\ Rev.\ \textbf{159}, 353 (1967).

\bibitem{ruby}
H.~K.~Mao, J.~Xu, and P.~M.~Bell,
J.\ Geophys.\ Res.\ \textbf{91}, 4673 (1986).

\bibitem{Tanida2022}
H.~Tanida, H.~Matsuoka, K.~Mitsumoto, Y.~Muro, T.~Fukuhara, and H.~Harima,
J.\ Phys.\ Soc.\ Jpn.\ \textbf{91}, 013704 (2022).

\bibitem{Aoyama}
T.~Aoyama \textit{et al.}, Phys.\ Rev.\ B, submitted.

\bibitem{Kawamura2022}
Y.~Kawamura, K.~Ikeda, A.~N.~B.~A.~Dalan, J.~Hayashi, K.~Takeda, C.~Sekine, T.~Matsumura, J.~Gouchi, Y.~Uwatoko, T.~Tomita,
H.~Takahashi, and H.~Tanida,
J. Phys. Soc. Jpn. \textbf{91}, 064714 (2022).

\bibitem{Fetisov1985}
E.~P.~Fetisov and D.~I.~Khomskii,
Solid State Commun. \textbf{56}, 403 (1985).

\bibitem{Hackl2008}
A.~Hackl and M.~Vojta,
Phys. Rev. B \textbf{77}, 134439 (2008).

\bibitem{deVisser1989}
A.~de~Visser, A.~Lacerda, P.~Haen, J.~Flouquet, F.~E.~Kayzel, and J.~J.~Franse,
Phys. Rev. B \textbf{39}, 11301 (1989).

\bibitem{Takeuchi2014}
T.~Takeuchi, S.~Yoshiuchi, Y.~Hirose, F.~Honda, R.~Settai, and Y.~\=Onuki,
JPS Conf. Proc. \textbf{3}, 011017 (2014).

\bibitem{CMS2024}
H.~Tanida, Y.~Yanagi, T.~Yamada, K.~Mitsumoto, N.~Higa, and T.~Matsumura,
J. Phys. Soc. Jpn. \textbf{93}, 073703 (2024).

\bibitem{Varma2016}
C.~M.~Varma,
Rep.\ Prog.\ Phys.\ \textbf{79}, 082501 (2016).

\bibitem{Welter1994}
R.~Welter, G.~Venturini, and B.~Malaman,
J.\ Alloys Compd.\ \textbf{206}, 55 (1994).

\bibitem{Sakai2026}
Y.~Sakai, F.~Hori, H.~Matsumura, S.~Oguchi, S.~Kitagawa, K.~Ishida, and H.~Tanida,
J. Phys. Soc. Jpn. \textbf{95}, 024702 (2026).

\bibitem{Chevalier2004}
B.~Chevalier and S.~F.~Matar,
Phys.\ Rev.\ B \textbf{70}, 174408 (2004).

\bibitem{Zhang2025}
S.-K.~Zhang, Y.~Xu, G.~Li, J.~Wang, Z.~Zhou, and Y.~An,
Phys. Rev. B \textbf{112}, 075142 (2025).

\end{thebibliography}
\end{document}